\documentclass[aps,prd,twocolumn,superscriptaddress,showpacs, 10pt]{revtex4}
\usepackage{amsmath,amssymb}
\usepackage{graphicx}
\usepackage{color}
\usepackage{caption}
\usepackage{hyperref}
\usepackage{graphicx}
\usepackage[brazilian]{babel}
\usepackage[utf8]{inputenc}
\usepackage[T1]{fontenc}
\usepackage[symbol]{footmisc}

\begin{document}

\title{Spin-polarized current, spin-transfer torque and spin Hall effect in presence of an electromagnetic non-minimal coupling}
\author{Rodrigo Turcati}\email{rodrigo.turcati@posgrad.ufsc.br}
\affiliation{Departamento de Física, Universidade Federal de Santa Catarina,
CEP 88040-900, Florianópolis, Santa Catarina, Brasil}
\author{Carlos Andres Bonilla Quintero}\email{bonilla@on.br}
\affiliation{Observatório Nacional, CEP 20921-400, Rio de Janeiro, RJ, Brasil}
\author{Jos\'{e} Abdalla Helay\"{e}l-Neto}\email{helayel@cbpf.br}
\affiliation{Departamento de Astrofísica, Cosmologia e Interações Fundamentais, Centro Brasileiro de Pesquisas F\'{i}sicas, Rua Dr. Xavier Sigaud 150, Urca, 22290-180, Rio de Janeiro, RJ, Brazil}
\author{Enrique Arias}\email{earias@iprj.uerj.br}
\affiliation{Instituto Politécnico, Universidade do Estado do Rio de Janeiro, 28625-570 Nova Friburgo, Brazil}

\def\d{{\mathrm{d}}}
\newcommand{\scri}{\mathscr{I}}
\newcommand{\sun}{\ensuremath{\odot}}
\def\J{{\mathscr{J}}}
\def\L{{\mathscr{L}}}
\def\sech{{\mathrm{sech}}}
\def\T{{\mathcal{T}}}
\def\tr{{\mathrm{tr}}}
\def\diag{{\mathrm{diag}}}
\def\ln{{\mathrm{ln}}}
\def\Horava{Ho\v{r}ava}
\def\Aether{\AE{}ther}
\def\AEther{\AE{}ther}
\def\aether{\ae{}ther}
\def\UH{{\text{\sc uh}}} 
\def\KH{{\text{\sc kh}}}

\begin{abstract}

In this contribution, we start off from a fully relativistic description of a single electron non-minimally coupled to an external electromagnetic field. Making direct use of the field equation, instead of canonically deriving from the Lagrangian density, the relativistic total angular momentum is attained, where the effect of the relativistic torque due to the external sources is also taken into account. Both the spin density and the orbital angular momentum tensors are identified in the relativistic-covariant approach. Furthermore, the symmetric and gauge invariant energy-momentum tensor is derived as well. In addition, to inspect the non-relativistic regime, a perturbative expansion of the field equation up to the leading order in $\left(v/c\right)$ is carried out, where spin-orbit interaction terms naturally emerge in the non-relativistic Hamiltonian as a consequence of the non-minimal coupling. Features related to the spin currents, the spin-transfer torque and their dependence on both magnetic and electric external fields are then calculated and discussed. Considering that spin-orbit coupled systems are of particular interest in the study of spin Hall effect, a two-dimensional (planar) scenario of the system is contemplated, where the contributions to the Landau levels are re-assessed. As a consequence of the planar regime and the non-minimal coupling, a peculiar sort of fractionalization of the spin-up and -down components emerges with the different spin components localized in distinct positions.

\end{abstract}

\maketitle
\section{Introduction}

Spin transport electronics, or spintronics for short, is a spin-based technology in which the intrinsic spin of the electron is, in addition to its electric charge, further explored in the study of transport phenomena in Condensed Matter Physics \cite{Prinz1998,wolf2001,Awschalom2002,Zutic2004}. The first ideas concerning the use of the spin as a degree of freedom in electronic advices date back to 1970s, where the anomalous Hall effect and the Mott scattering phenomena were considered to theoretically forecast the extrinsic spin Hall effect \cite{Dyakonov1971,Dyakonov71,Mott1929,Mott1965}. More recently, in Refs. \cite{Slonczewski1995,Berger1996}, it was suggested that a spin-polarized current may induce magnetic switching and dynamic excitations in ferromagnet spin systems. Ever since, an outstanding progress in both theoretical and experimental features concerning spintronics have been achieved, rendering feasible a wide variety of applications \cite{Hirsch1999,Zhang2000,Kato2004,Wunderlich2005,Saitoh2006,Valenzuela2006,Zhao2006}.

Currently, there has been a growing interest in the attempt to understand features related to spin-orbit coupled systems \cite{Muramaki2003,Sinova2004}. 
Although the spin-orbit coupling is measured in the realm of the nonrelativistic quantum mechanics, this interaction indeed is a relativistic correction that arises from a fully relativistic description of the electron. In this sense, the spin-orbit coupling has an intrinsic relation with the relativistic quantum nature of the spin, which is responsible for several interesting quantum phenomena, such as the spin precession \cite{Datta1990,Matsuyama} and the spin Hall effect \cite{Sinova2006,Engel2007}, for instance. 

The understanding of the quantum nature of the electron spin dates back to the famous Stern--Gerlach experiment \cite{Gerlach}, where measurements of electron magnetic dipole moment in presence of an inhomogeneous magnetic field demonstrated the existence of an intrinsic spin angular momentum. To account for the interaction of the electron spin with an external magnetic field in the framework of non-relativistic quantum mechanics, Pauli formulated Schrödinger equation by introducing two-component (non-relativistic) spinors and the Clifford algebra of three-dimensional space and modified the kinetic term by introducing the Pauli matrices, i.e., $\boldsymbol{p^{2}\rightarrow(\sigma\cdot{p})(\sigma\cdot{p})}$ \cite{Pauli:1927}. In the absence of interactions, there is no difference with respect the Schrödinger equation. On the other hand, when one takes the electromagnetic minimal coupling, $p^{\mu}\rightarrow{p^{\mu}-eA^{\mu}}$ into account, it leads to the appearance of a spin-dependent interaction Hamiltonian 
\begin{equation}\label{Zeemanterm}
\mathcal{H}_{I}=-\boldsymbol{\mu}\cdot\boldsymbol{B}, 
\end{equation}
where $\boldsymbol{\mu}\left(=\frac{e\hbar}{2mc}\boldsymbol{\sigma}\right)$ is the electron magnetic dipole moment and $\boldsymbol{B}$ is the external magnetic field. The prescription proposed by Pauli was sucessful to ensure the right interaction between the electron spin and the magnetic field. Nevertheless, one year later, Dirac succeeded in proposing a fully relativistic quantum description of the electron \cite{Dirac:1928hu}, given by the equation 
\begin{equation}\label{de}
\left(i\hbar\gamma^{\mu}\partial_{\mu}-mc-\frac{e}{c}\gamma^{\mu}A_{\mu}\right)\psi=0, 
\end{equation}
where $m$ is the electron mass, $A^{\mu}$ is the external electromagnetic four-potential subject to a $U(1)$-transformation and $\gamma$ refers to the Dirac gamma-matrices. Indeed, when one goes over into the low-relativistic limit of Eq. (\ref{de}), the Hamiltonian term (\ref{Zeemanterm}) automatically appears in the Schrodinger wave equation, showing that the relativistic Dirac equation can naturally accommodate the spin-$1/2$ nature of the electron without any {\it ad hoc} assumptions. Furthermore, the spin magnetic moment generated in this framework provides the correct gyromagnetic ratio, $g=2$, for the electron. 

Besides its undoubtful success, the Dirac equation does not provide a complete description for spin-$1/2$ particles at very high energies. The reason is due to corrections to observable quantities that only Quantum Electrodynamics is able to account for through phenomena such as vacuum polarization and pair production, for instance. Besides that, experimental measurements concerning the gyromagnetic ratio, $g$, of the electron reveal that its value was not exactly $g=2$ \cite{Kusch:1948mvb}, as predicted by the Dirac equation. The so-called {\it anomalous magnetic moment of the electron} (and for the muon and tau leptons) is  correctly reproduced by Quantum Electrodynamics and, more generally, by the Standard Model. One of the first attempts to solve the mentioned problem was due to Pauli, who suggested that, besides the electromagnetic minimal coupling, one should also take a coupling directly with the Maxwell tensor $F_{\mu\nu}$. Later on, the advent of the Quantum Electrodynamics provided the correct description of the aforementioned problems \cite{Schwinger:1948iu}. Nevertheless, the Pauli anomalous moment interaction is still used to describe several physical phenomena such as the Aharonov-Casher effect for neutral particles \cite{Aharonov:1984xb}, among others \cite{Anandan:1989lvk,Sastry:1999is,Bakke:2008xt,Bakke:2008yk,Bakke:2012wu,Teruel:2014iza,Hassanabadi:2016jtt}. Indeed, this coupling may be interpreted as an effective interaction of fermions with a non-zero magnetic dipole moment. 

In the attempt to further extend ideas related to electron spin beyond the electromagnetic minimal coupling, we here discuss a fully relativistic description of the spin in presence of the Pauli interaction and investigate the consequences at the non-relativistic limit. In particular, spin-orbit coupling terms come out. Afterwards, the dynamics of spin currents is investigated, as well as modifications in the charge continuity equation associated to the local $U(1)$ gauge symmetry. We also comment on the spin Hall current. To conclude, contributions to the quantum Landau levels are derived and a sort of fractionalization between the spin-up and -down electron states is observed.

The structure of this paper is organized as follows. We start by introducing and briefly re-deriving the Dirac equation in the presence of both minimal and non-minimal electromagnetic couplings. This is done in Sec. \ref{review}. The non-relativistic regime is described in Sec. \ref{nonrelativisticsection}. In Sec. \ref{Noethersection}, we evaluate the influence of the Pauli interaction on the charge density current. In Sec. \ref{angularmomentumsection}, the relativistic total angular momentum and the time evolution of the spin density tensor, along with the stress-energy tensor, are derived. The rôle of the relativistic torque is investigated. Sec. \ref{spindensitysection} is devoted to the discussion of the spin currents. Features related to the spin Hall effect, as well as aspects of Landau levels in our study are discussed in Sec. \ref{spindensitysection}. Our Conclusions and Future Perspectives are presented in Sec. \ref{conclusion}.

We shall adopt the Heaviside-Lorentz units system. In our conventions, the signature of the Minkowski metric is $(+,-,-,-)$.


\section{Electromagnetic non-minimally coupled Dirac equation}\label{review}

In this Section, basic features of the relativistic Dirac equation minimally and non-minimally coupled to the electromagnetic field are briefly reviewed.  We start off with the relativistic wave equation for the massive spinor field $\psi$, namely, 
\begin{equation}
\left(i\hbar\gamma^{\mu}\partial_{\mu}-mc\right)\psi=0.
\end{equation}

As it is well-known, the relativistic description of the electron is fully accomplished when the Dirac field is coupled to the gauge potential, $A^{\mu}$. This is the so-called minimal prescription, and it is implemented by means of the covariant derivative 
\begin{equation}
\partial_{\mu}\rightarrow D_{\mu}\equiv\partial_{\mu}+\frac{ie}{\hbar{c}}A_{\mu},
\end{equation}
and by the local phase transformation of the spinor field, ${\psi}'=e^{-\frac{ie}{\hbar{c}}\alpha(x)}{\psi}$, where $\alpha(x)$ is the local gauge parameter related to the $U(1)$-symmetry group. Although the minimal coupling is commonly used to describe electrically charged particles, there are other gauge interactions that can be introduced into the electron dynamical equation and still preserve the gauge symmetry. Thus, besides the gauge field $A^{\mu}$, one may also couple the Dirac particle to the Maxwell tensor $F^{\mu\nu}$. One possible way to introduce an electromagnetic non-minimal coupling is through the so-called Pauli interaction, which reads as $\mathcal{L}_{Pauli}=-\left(e\hbar{\kappa}/8mc^{2}\right)\bar{\psi}\sigma^{\mu\nu}\psi{F}_{\mu\nu}$. 

The Lagrangian that describes the Dirac field minimally and non-minimally coupled to the electromagnetic field is 
\begin{eqnarray}\label{PauliLagrangian}
\mathcal{L}\!\!&=&\!\!\frac{i\hbar}{2}\!\left[\bar{\psi}\left(\gamma^{\mu}\partial_{\mu}\psi\right)\!-\!\left(\partial_{\mu}\bar{\psi}\right)\gamma^{\mu}\psi\right]\!-\!mc\bar{\psi}\psi\!-\!\frac{e}{c}\bar{\psi}\gamma^{\mu}\psi{A_{\mu}}\nonumber\\
&&-\frac{e\hbar{\kappa}}{8mc^{2}}\bar{\psi}\sigma^{\mu\nu}\psi{F}_{\mu\nu}, 
\end{eqnarray}
where the matrix $\sigma^{\mu\nu}$ is defined as
\begin{eqnarray}
\sigma^{\mu\nu}&=&\frac{i}{2}\left[\gamma^{\mu},\gamma^{\nu}\right],
\end{eqnarray}
and $\kappa$ is a dimensionless parameter measuring the strength of the electron anomalous magnetic moment contribution.

The corresponding field equation for the Dirac spinor $\psi$ is
\begin{eqnarray}\label{Diracequation}
\left(i\hbar\gamma^{\mu}D_{\mu}-mc-\frac{e\hbar{\kappa}}{8mc^{2}}\sigma^{\mu\nu}{F}_{\mu\nu}\right)\psi=0. 
\end{eqnarray}

We should highlight the reason why we introduce the electromagnetic non-minimal coupling. Ref. \cite{mondal2018} investigates and derives the spin-transfer and the spin-orbit torques from the Dirac equation in its relativistic regime by considering minimal electromagnetic coupling. In our contribution, we also inspect the same issues but we introduce from the very onset the non-minimal coupling as shown in Eq. (\ref{Diracequation}). The inclusion of the non-minimal term  brings about new effects involving the electric field. This point shall become more evident when we will be commenting below Eq. (\ref{paulitypeequaiton}) of Sec. (\ref{nonrelativisticsection}) on the specific contributions arising from the non-minimal coupling which involve the purely electric sector.

We also motivate the inclusion of the non-minimal coupling based on arguments of effective quantum field theories. Ultraviolet finite radiative corrections in QED naturally induce this non-minimal coupling; therefore, its inclusion corresponds to taking into account a (one-loop) contribution that  affects the electric sector with new terms that depende on the electric field and corrects the value $g=2$ of the electron gyromagnetic ratio by including a field-theoretic effect.

One readily derives the Hamiltonian that arises from Eq. (\ref{Diracequation}), namely,
\begin{eqnarray}\label{diracHamiltonian}
\mathcal{H}_{D}\!\!&=&\!\!c\boldsymbol{\alpha}\cdot\boldsymbol{\Pi}+\beta{mc^{2}}+e\phi+i\frac{e\hbar{\kappa}}{4mc}\beta\boldsymbol{\alpha}\cdot\boldsymbol{E}-\frac{e\hbar{\kappa}}{4mc}\beta\boldsymbol{\Sigma}\cdot\boldsymbol{B}, \nonumber\\
\end{eqnarray}
where $\boldsymbol{\Pi}=-i\hbar\boldsymbol{\nabla}-\frac{e}{c}\boldsymbol{A}$ is the generalized kinetic operator, $\phi$ and $\boldsymbol{A}$ are the electric and magnetic vector potentials, respectively. In addition, we adopt the standard Dirac $\gamma$-matrices representation, i.e.,
\begin{equation}
\gamma^{0}=\left[\begin{array}{cc}
1 & 0 \\ 0 & -1  
\end{array}\right] \gamma^{i}=\left[\begin{array}{cc}
0 & \sigma^{i} \\ -\sigma^{i} & 0  
\end{array}\right] \gamma^{5}=\left[\begin{array}{cc}
0 & 1 \\ 1 & 0  
\end{array}\right]\nonumber
\end{equation}
together with the following definitions
\begin{equation}\label{definitions}
\beta=\gamma^{0}, \quad\quad \alpha^{i}=\gamma^{0}\gamma^{i} \quad\textnormal{and}\quad \Sigma^{\mu}=\gamma^{0}\gamma^{\mu}\gamma^{5}. 
\end{equation}

Notice that the first three terms on the rhs of (\ref{diracHamiltonian}) represent the standard Dirac Hamiltonian for the electron. Besides that, there is the presence of two additional terms at the rhs of the Hamiltonian (\ref{diracHamiltonian}), which stem from the electromagnetic non-minimal coupling. Indeed, unlike the minimal coupling case, the Pauli interaction induces the appearance of both electric and magnetic fields, instead of the potentials, into the Dirac Hamiltonian.


\section{Non-relativistic Pauli-Schrödinger equation and the spin-orbit interaction}\label{nonrelativisticsection}

Several phenomena, such as the spin Hall effect, the Landau-Lifshitz-Gilbert equation, among others, are fairly-well described at the domain of the non-relativistic quantum mechanics. In this sense, it would be instructive to investigate what sort of effects may emerge in the dynamics of the electron at the low-relativistic approximation whenever the non-minimal (Pauli) interaction is included. 

In order to obtain the non-relativistic limit of the Dirac Hamiltonian (\ref{diracHamiltonian}), one initially splits the spinor $\psi$ in its relativistic $\xi$ and non-relativisticic $\varphi$ components, i.e., $\psi=\begin{pmatrix}
    \varphi  \\
    \xi
\end{pmatrix}
$. Inserting the mentioned decomposition in Eq. (\ref{Diracequation}), the equations for $\xi$ and $\phi$ are, in momentum space, respectively, 
\begin{eqnarray}\label{non-relativisticiccomponent}
&&\left[E-e\phi-mc^{2}+\frac{e\hbar\kappa}{4mc}\boldsymbol{\sigma}\cdot\boldsymbol{B}\right]\varphi=\nonumber\\
&&=c\boldsymbol{\sigma}\cdot\left(-i\hbar\boldsymbol{\nabla}-\frac{e}{c}\boldsymbol{A}+\frac{ie\hbar\kappa}{4mc^{2}}\boldsymbol{E}\right)\xi, \\ \label{relativisticcomponent}
&&\left[E-e\phi+mc^{2}-\frac{e\hbar\kappa}{4mc}\boldsymbol{\sigma}\cdot\boldsymbol{B}\right]\xi=\nonumber\\
&&=c\boldsymbol{\sigma}\cdot\left(-i\hbar\boldsymbol{\nabla}-\frac{e}{c}\boldsymbol{A}-\frac{ie\hbar\kappa}{4mc^{2}}\boldsymbol{E}\right)\varphi. 
\end{eqnarray}

Let us now decouple the weakly-relativistic component $\varphi$ from the above set of equations. If one takes the non-relativisticic limit, i.e., the regime where $v/c\ll1$, the rest-frame energy reduces to $E\approx{mc^{2}}$, and the interaction terms may be neglected on the lhs of Eq. (\ref{relativisticcomponent}). Under these approximations, Eq. (\ref{relativisticcomponent}) is then as follows: 
\begin{equation}\label{xicomponent}
\xi\approx\frac{1}{2mc}\boldsymbol{\sigma}\cdot\left(-i\hbar\boldsymbol{\nabla}-\frac{e}{c}\boldsymbol{A}-\frac{ie\hbar\kappa}{4mc^{2}}\boldsymbol{E}\right)\varphi. 
\end{equation}

Inserting Eq. (\ref{xicomponent}) into the Eq. (\ref{non-relativisticiccomponent}), one promptly finds the non-relativisticic gauge invariant Hamiltonian
\begin{eqnarray}\label{paulitypeequaiton}
H\!&=&\!\frac{1}{2m}\left(-i\hbar\boldsymbol{\nabla}-\frac{e}{c}\boldsymbol{A}+\frac{e\hbar{\kappa}}{4mc^{2}}\boldsymbol{E}\times\boldsymbol{\sigma}\right)^{2}-\boldsymbol{\mu}_{eff}\cdot\boldsymbol{B}\nonumber\\
&&+e\phi-\frac{e^{2}\hbar^{2}{\kappa^{2}}}{32m^{3}c^{4}}\boldsymbol{E}^{2}-\frac{e\hbar^{2}\kappa}{8m^{2}c^{2}}\boldsymbol{\nabla}\cdot\boldsymbol{E}, 
\end{eqnarray}
where $\boldsymbol{\mu}_{eff}\equiv\frac{e\hbar}{2mc}\left(\frac{2+\kappa}{2}\right)\boldsymbol{\sigma}$ is the electron magnetic dipole moment corrected by the Pauli coupling. 

Now, let us return to Eq. (\ref{paulitypeequaiton}). As far as the magnetic dipole moment is concerned, one promptly notices that $\boldsymbol{\mu}$ receives a small contribution that depends on the strength of the $\kappa$-parameter. It should not come as a surprise, since the Pauli non-miminal coupling was historically introduced to explain the deviation from the theoretical prevision of the electron gyromagnetic factor $g=2$. In addition, there is a quadratic electric-field-dependent term $\left(e^{2}\hbar^{2}{\kappa^{2}}/32m^{3}c^{4}\right)\boldsymbol{E}^{2}$. The latter is related to an inhomogeneous background charge distribution. Furthermore, the generalized momentum is modified due the appearance of $\left(e\hbar{\kappa}/4mc^{2}\right)\boldsymbol{E}\times\boldsymbol{\sigma}$. In fact, such term denotes the presence of a geometrical phase due to the electric field similar to the Aharonov-Casher effect for neutral particles \cite{Aharonov:1984xb}. 

Let us then rewrite the above Hamiltonian in order to get a better understanding of these new terms. The Hermitian Pauli-type Hamiltonian (\ref{paulitypeequaiton}) can conveniently be cast as
\begin{eqnarray}\label{so}
H\!&=&\!\frac{\boldsymbol{\Pi}^{2}}{2m}-\boldsymbol{\mu}_{eff}\cdot\boldsymbol{B}+e\phi+\frac{e^{2}\hbar^{2}{\kappa^{2}}}{32m^{3}c^{4}}\boldsymbol{E}^{2}-\frac{e\hbar^{2}\kappa}{8m^{2}c^{2}}\boldsymbol{\nabla}\cdot\boldsymbol{E}\nonumber\\
&&-\frac{ie\hbar^{2}\kappa}{8m^{2}c^{2}}\boldsymbol{\sigma}\cdot\left(\boldsymbol{\nabla}\times\boldsymbol{E}\right)-\frac{e\hbar\kappa}{4m^{2}c^{2}}\boldsymbol{\sigma}\cdot\left(\boldsymbol{E}\times\boldsymbol{\Pi}\right).
\end{eqnarray}

Comparing with the Hamiltonian (\ref{paulitypeequaiton}), there appear two new terms. Indeed, the last two terms at the rhs of (\ref{so}) denote spin-dependent interactions terms due to the non-miminal coupling. Actually, these terms describe the full spin-orbit coupling, i.e., the particle momentum coupled to its spin. In addition, the anomalous velocity operator is $\boldsymbol{v}=(1/m)\boldsymbol{\Pi}+(e\hbar\kappa/4m^{2}c^{2})\boldsymbol{E}\times\boldsymbol{\sigma}$.

As an aside comment, we would also like to remark that the spin-orbit coupling is absent in the Pauli-Schrödinger equation. To obtain the spin-orbit interaction in the non-relativisticic Hamiltonian, one needs to expand the Dirac equation up to the order $\left(v/c\right)^{2}$ in the perturbative formalism. This is in contrast with our model, where the spin-orbit coupling term naturally gives rise at the first-order level in the perturbative approach; this is so by virtue of the non-minimal coupling. 


\section{Conserved Noether currents}\label{Noethersection}

Let us now focus on the electric current in the non-relativisticic regime. To begin with, it is important to notice that the infinitesimal phase variation $\delta\psi=-\left(ie/\hbar{c}\right)\alpha(x)\psi$, where $\alpha(x)$ is the parameter related to the $U(1)$ local gauge symmetry, together with the transformation of the fields $A'_{\mu}=A_{\mu}+\partial_{\mu}\alpha(x)$, lead the Lagrangian (\ref{PauliLagrangian}) invariant under the gauge symmetry. According to the Noether's theorem, there must exist a conserved current $j^{\mu}=\left({c}\rho,\boldsymbol{j}\right)$ related to such symmetry, namely,
\begin{eqnarray}
j^{\mu}=e\bar{\psi}\gamma^{\mu}\psi, 
\end{eqnarray}
which satisfies the continuity equation $\partial_{\mu}j^{\mu}=0$.

By employing the decomposition of $\psi$ in its relativistic, $\xi$, and non-relativistic, $\varphi$, components into the four-vector current, $j^{\mu}$, the charge density $\rho$ and the spatial density current $\boldsymbol{j}$, up to the leading order in $v/c$, are, respectively,
\begin{eqnarray}
\rho=e\varphi^{\dagger}\varphi
\end{eqnarray}
and 
\begin{eqnarray}\label{electricdensitycurrent}
\boldsymbol{j}&=&\frac{ie\hbar}{2m}\left[\left(\boldsymbol{\nabla}\varphi^{\dagger}\right)\varphi-\varphi^{\dagger}\left(\boldsymbol{\nabla}\varphi\right)\right]-\frac{e^{2}}{mc}\left(\varphi^{\dagger}\boldsymbol{A}\varphi\right)\nonumber\\
&&+\frac{e\hbar}{2m}\boldsymbol{\nabla}\times\left(\varphi^{\dagger}\boldsymbol{\sigma}\varphi\right)+\frac{e^{2}\hbar\kappa}{4m^{2}c^{2}}\left(\boldsymbol{E}\times\varphi^{\dagger}\boldsymbol{\sigma}\varphi\right).
\end{eqnarray}

Comparing with the conventional electric current, that is related to the flow of slowly moving charges, the density current (\ref{electricdensitycurrent}) receives a contribution from the spin-orbit interaction. Indeed, an external electric field will induce a spin-dependent contribution to the electric current, a term that describes the so-called Anomalous Hall Effect. This phenomenon is commonly observed in ferromagnetic and non-magnetic conductors. 

To conclude, we would like to remark that the charge current is conserved even in the presence of the Pauli interaction. It is an expected result since the non-minimally coupling maintain the gauge symmetry. 


\section{The gauge-invariant energy-momentum tensor and the relativistic total angular momentum}\label{angularmomentumsection}

One of our purposes of this contribution is to discuss the relativistic description of the electron spin in the electromagnetic non-miminal scenario. Such a description can be achieved by first analyzing the total angular momentum of the Lagrangian (\ref{PauliLagrangian}). To go further with our program, let us initially exploit the invariance of the Lagrangian under Lorentz rotations. Under infinitesimal Lorentz transformations, the fields transform as below:
\begin{eqnarray}
\delta\psi\!=\!-\frac{i}{2}w^{\alpha\beta}\Sigma_{\alpha\beta}\psi, \!\!\quad\!\! \delta\bar{\psi}\!=\!\frac{i}{2}\bar{\psi}\Sigma_{\alpha\beta}w^{\alpha\beta}, \!\!\quad\!\! \delta{A^{\mu}}\!=\!w^{\mu\nu}A_{\nu},\nonumber\\
\end{eqnarray}
where $w^{\alpha\beta}\left(=-w^{\beta\alpha}\right)$ is a constant parameter of the Lorentz transformations and $\Sigma^{\alpha\beta}\left(=\sigma^{\alpha\beta}/2\right)$ are the Lorentz group generators in the representation where the spinor fields sit. Furthermore, the space-time coordinates undergo the infinitesimal transformation $\delta{x^{\mu}}=w^{\mu\nu}x_{\nu}$. 

Noether's theorem, in turn, ensures the existence of a conserved current, $\mathcal{J}^{\mu\alpha\beta}\left(=\mathcal{L}^{\mu\alpha\beta}+\mathcal{S}^{\mu\alpha\beta}\right)$, as a consequence of Lorentz symmetry, where
\begin{eqnarray}
\mathcal{L}^{\mu\alpha\beta}&=&T_{D}^{\mu\alpha}x^{\beta}-T_{D}^{\mu\beta}x^{\alpha}\nonumber\\
&&-\frac{e\hbar\kappa}{4mc^{2}}\left(\bar{\psi}\sigma^{\mu\lambda}\psi\right)\left(x^{\alpha}\partial^{\beta}-x^{\beta}\partial^{\alpha}\right)A_{\lambda}
\end{eqnarray} 
is the orbital angular momentum and 
\begin{eqnarray}
\mathcal{S}^{\mu\alpha\beta}=\frac{\hbar}{4}\bar{\psi}\{\gamma^{\mu},\sigma^{\alpha\beta}\}\psi-\frac{e\hbar\kappa}{4mc^{2}}\bar{\psi}\left(\sigma^{\mu\alpha}A^{\beta}-\sigma^{\mu\beta}A^{\alpha}\right)\psi\nonumber\\
\end{eqnarray} 
is the spin density tensor. 

Furthermore, $T^{\mu\nu}_{D}$ is the standard Dirac energy-momentum tensor\footnotemark[1]\footnotetext[1]{Notice that $T^{\mu\nu}_{D}$ is not the energy-momentum tensor that would arise from the invariance under space-time translation of the Lagrangian. The canonical stress tensor corresponding to the Lagrangian (\ref{PauliLagrangian}) is as follows:  
$$T^{\mu\nu}=\frac{i\hbar}{2}\left[\bar{\psi}\gamma^{\mu}\left(\partial^{\nu}\psi\right)-\left(\partial^{\nu}\bar{\psi}\right)\gamma^{\mu}\psi\right]-\frac{e\hbar\kappa}{4mc^{2}}\left(\bar{\psi}\sigma^{\mu\nu}\psi\right)-\eta^{\mu\nu}\mathcal{L}.$$}, given by 
\begin{eqnarray}
T_{D}^{\mu\nu}=\frac{i\hbar}{2}\left[\bar{\psi}\gamma^{\mu}\left(\partial^{\nu}\psi\right)-\left(\partial^{\nu}\bar{\psi}\right)\gamma^{\mu}\psi\right]-\eta^{\mu\nu}\mathcal{L}_{D},
%
%
\end{eqnarray}
where $\mathcal{L}_{D}$ is the Dirac Lagrangian minimally coupled to the electromagnetic field.

Although the total canonical angular momentum is conserved, it is not gauge-invariant. The gauge dependence on the angular momentum  prevents such this quantity to be associated to an observable. Indeed, the standard prescription for computing this tensor suffers the drawback of being neither symmetric, not gauge-invariant. On the other hand, there are many attempts to improve both energy-momentum and angular momentum density tensors. To circumvent the gauge-dependence problem, let us then follow another path to obtain the total angular momentum.   

The algorithm to derive the total angular momentum consists in starting off from the field equations rather than considering the Lagrangian density. To do this, we first consider the Lorentz transformation of the spinor field, i.e., 
\begin{eqnarray}\label{fieldvariation}
\delta_{L}\psi&=&\frac{1}{2}w^{\alpha\beta}\left[\left(x_{\alpha}D_{\beta}-x_{\beta}D_{\alpha}\right)-i\Sigma_{\alpha\beta}\right]\psi, \\
\label{adjointfieldvariation}\delta_{L}\bar{\psi}&=&\frac{1}{2}w^{\alpha\beta}\bar{\psi}\left[\left(x_{\alpha}\overleftarrow{D}^{\dagger}_{\beta}-x_{\beta}\overleftarrow{D}^{\dagger}_{\alpha}\right)+i\Sigma_{\alpha\beta}\right], 
\end{eqnarray}
where, in our prescription, the ordinary derivative is modified by the gauge covariant derivative $\left(\partial_{\mu}\rightarrow{D_{\mu}}\equiv\partial_{\mu}+\left(ie/\hbar{c}\right)A_{\mu}\right)$ in order to ensure the gauge invariance of the total angular momentum. 

Next, we couple the field equation (\ref{Diracequation}) and its Dirac-adjoint to the field variations (\ref{fieldvariation}) and (\ref{adjointfieldvariation}), which yields
\begin{eqnarray}
&&\delta_{L}\bar{\psi}\left[i\hbar\gamma^{\mu}D_{\mu}-mc-\frac{e\hbar\kappa}{8mc^{2}}\sigma^{\mu\nu}F_{\mu\nu}\right]\psi\nonumber\\
&&-\left[i{\hbar}\left(D^{\dagger}_{\mu}\bar{\psi}\right)\gamma^{\mu}+\bar{\psi}mc+\frac{e\hbar\kappa}{8mc^{2}}\bar{\psi}\sigma^{\mu\nu}F_{\mu\nu}\right]\delta_{L}\psi=0.\nonumber\\
\end{eqnarray}

After some steps of algebraic manipulations, the equation for the total angular momentum $\mathcal{J}^{\mu\alpha\beta}$ turns out to be
\begin{eqnarray}\label{angularmomentumequation}
\partial_{\mu}\mathcal{J}^{\mu}_{\phantom{\mu}\alpha\beta}&=&\mathcal{T}_{\alpha\beta},
\end{eqnarray}
where $\mathcal{T}_{\alpha\beta}$ is the relativistic torque, which is given by
\begin{eqnarray}\label{relativistictorque}
\mathcal{T}_{\alpha\beta}&=&\frac{j^{\mu}}{c}\left(x_{\alpha}F_{\mu\beta}-x_{\beta}F_{\mu\alpha}\right)\nonumber\\
&&-\frac{e\hbar\kappa}{8mc^{2}}\left(\bar{\psi}\sigma^{\mu\nu}\psi\right)\left(x_{\alpha}\partial_{\beta}-x_{\beta}\partial_{\alpha}\right)F_{\mu\nu}\nonumber\\
&&-\frac{e\hbar\kappa}{4mc^{2}}\bar{\psi}\left(\sigma^{\mu}_{\phantom{\mu}\alpha}F_{\mu\beta}-\sigma^{\mu}_{\phantom{\mu}\beta}F_{\mu\alpha}\right)\psi.
\end{eqnarray}

Now, the orbital angular momentum takes the form
\begin{eqnarray}
\mathcal{L}^{\mu}_{\phantom{\mu}\alpha\beta}&=&\theta^{\mu}_{\phantom{\mu}\alpha}x_{\beta}-\theta^{\mu}_{\phantom{\mu}\beta}x_{\alpha}
\end{eqnarray} 
and the spin density tensor reduces to
\begin{eqnarray}
\mathcal{S}^{\mu}_{\phantom{\mu}\alpha\beta}=-\frac{\hbar}{4}\bar{\psi}\{\gamma^{\mu},\sigma_{\alpha\beta}\}\psi.
\end{eqnarray} 
Notice that both the orbital and the spin density components are gauge-symmetric, as a physically realizable quantity should be.

In addition, the gauge-invariant energy-momentum tensor takes the form below:
\begin{eqnarray}
\theta^{\mu\nu}=\frac{i\hbar}{2}\left[\bar{\psi}\gamma^{\mu}\left(D^{\nu}\psi\right)-\left(D^{\nu}\bar{\psi}\right)\gamma^{\mu}\psi\right]-\eta^{\mu\nu}\mathcal{L}.
\end{eqnarray}

It is important to point out that the stress-energy tensor cast above is obtained by considering the same procedure that was adopted to derive the total angular momentum $\mathcal{J}^{\mu\alpha\beta}$, namely, by coupling the gauge variation to the field equations. In fact, to derive the energy-momentum, one needs to multiply the field equations by the field variations under space-time translations, i.e., $\delta\psi=-a^{\mu}D_{\mu}$, where $a^{\mu}$ is an arbitrary four-vector and $D_{\mu}$ is the covariant derivative. 

As for the dynamics of the orbital momentum and spin density, both tensors fulfill the following gauge-invariant equations:
\begin{eqnarray}\label{orbitalangularmomentumequation}
\partial_{\mu}\mathcal{L}^{\mu}_{\phantom{\mu}\alpha\beta}&=&\theta_{\beta\alpha}-\theta_{\alpha\beta}+\frac{j^{\mu}}{c}\left(x_{\alpha}F_{\mu\beta}-x_{\beta}F_{\mu\alpha}\right)\nonumber\\
&&-\frac{e\hbar\kappa}{8mc^{2}}\left(\bar{\psi}\sigma^{\mu\nu}\psi\right)\left(x_{\alpha}\partial_{\beta}-x_{\beta}\partial_{\alpha}\right)F_{\mu\nu}
\end{eqnarray}
and
\begin{eqnarray}\label{spindensityequation}
\partial_{\mu}\mathcal{S}^{\mu}_{\phantom{\mu}\alpha\beta}&=&\theta_{\alpha\beta}-\theta_{\beta\alpha}-\frac{e\hbar\kappa}{4mc^{2}}\bar{\psi}\left(\sigma^{\mu}_{\phantom{\mu}\alpha}F_{\mu\beta}-\sigma^{\mu}_{\phantom{\mu}\beta}F_{\mu\alpha}\right)\psi,\nonumber\\
\end{eqnarray}
which are not separately conserved. 

Furthermore, the energy-momentum tensor obeys the equation
\begin{eqnarray}\label{divergencestresstensor}
\partial_{\mu}\theta^{\mu\nu}=\frac{1}{c}F^{\nu\mu}j_{\mu}+\frac{e\hbar\kappa}{8mc^{2}}\left(\bar{\psi}\sigma^{\alpha\beta}\psi\right)\partial^{\nu}F_{\alpha\beta}. 
\end{eqnarray}

In order to get a more accurate understanding of the terms presented in the relativistic torque (\ref{relativistictorque}), let us introduce the Maxwell field strength into the Lagrangian (\ref{PauliLagrangian}). 

Therefore, the field equations associated to the gauge field $A^{\mu}$ are
\begin{eqnarray}
\partial_{\mu}F^{\mu\nu}=\frac{j^{\nu}}{c}-\frac{e\hbar\kappa}{4mc^{2}}\partial_{\mu}\left(\bar{\psi}\sigma^{\mu\nu}\psi\right),
\end{eqnarray}
or, equivalently,
\begin{eqnarray}
\boldsymbol{\nabla}\cdot\boldsymbol{D}&=&\rho, \\
\boldsymbol{\nabla}\times\boldsymbol{H}-\frac{1}{c}\partial_{t}\boldsymbol{D}&=&\frac{\boldsymbol{j}}{c},
\end{eqnarray}
where $\boldsymbol{D}\equiv\boldsymbol{E}+\boldsymbol{P}$ is the electric displacement and $\boldsymbol{H}\equiv\boldsymbol{B}-\boldsymbol{M}$ is the auxiliary field. The set of equations above assume the same form of the Maxwell equations in presence of matter, where we here have identified  $\boldsymbol{P}=-\left(ie\hbar\kappa/4mc^{2}\right)\left(\psi^{\dagger}\beta\boldsymbol{\alpha}\psi\right)$ as the electric polarization tensor and $\boldsymbol{M}=\left(e\hbar\kappa/4mc^{2}\right)\left(\psi^{\dagger}\beta\boldsymbol{\Sigma}\psi\right)$ as the magnetization tensor of the electron. 

We return now to Eq. (\ref{angularmomentumequation}).  Considering the space components $\left(\alpha=i,\beta=j\right)$, the time evolution of the angular momentum component $J^{0}_{\phantom{0}ij}=-\epsilon_{ijk}J^{k}$, where $J^{k}$ is the standard total angular momentum, takes the form
\begin{eqnarray}
d_{t}\boldsymbol{J}&=&\boldsymbol{r}\times\boldsymbol{f}+cP_{i}\left(\boldsymbol{r}\times\boldsymbol{\nabla}\right)E_{i}+cM_{i}\left(\boldsymbol{r}\times\boldsymbol{\nabla}\right)B_{i}\nonumber\\
&&+c\boldsymbol{P}\times\boldsymbol{E}+c\boldsymbol{M}\times\boldsymbol{B},
\end{eqnarray}
where $\boldsymbol{f}=\rho\boldsymbol{E}+\boldsymbol{j}\times\boldsymbol{B}$ is the Lorentz force density.

At this point, some comments are in order.
\begin{enumerate}
 \item[(i)] In the absence of external sources, the relativistic torque (\ref{relativistictorque}) vanishes, and the total angular momentum $\mathcal{J}^{\alpha}_{\phantom{\alpha}\mu\nu}$ is conserved, as expected. Furthermore, the energy-momentum tensor is conserved as well. 
 
 \item[(ii)]The first term on the rhs of (\ref{relativistictorque}) is the torque due to the standard Lorentz force. The second is a consequence of the couple between the orbital and spin degrees of freedom. From Eq. (\ref{orbitalangularmomentumequation}), it is clear that both torques influence the the dynamics of the orbital angular momentum. 
 
 \item[(iii)] The third term on the rhs of (\ref{relativistictorque}) is a torque on the spin degrees of fredoom due to the external electromagnetic field on the spin density, which is felt by its polarization and magnetization vectors. 
 
 \item[(iv)] Besides the Lorentz force, the divergence of the stress-energy tensor (\ref{divergencestresstensor}) receives an additional contribution coming from the Pauli interaction.
 
\end{enumerate}

Let us now focus our attention on the spin density tensor. As it is well-known, the notion of spin is related to the purely spatial component of the spin tensor density, i.e., $\mathcal{S}^{0}_{\phantom{0}ij}=-\epsilon_{ijk}S^{k}$, where $S^{k}=\left(\hbar/2\right)\left(\psi^{\dagger}\Sigma^{k}\psi\right)$ is the relativistic spin operator. Such notion relies on the fact that the $\sigma^{ij}$ matrices are the generators of the $SO(3)$ rotation subgroup of Lorentz group.  

Now, for the time evolution of the spin density, we perform the spatial integration of $S^{\mu}_{\phantom{\mu}ij}$, to obtain
\begin{eqnarray}
&&\int{d^{3}x}\left[\partial_{0}\mathcal{S}^{0}_{\phantom{\mu}ij}+\partial_{k}\mathcal{S}^{k}_{\phantom{\mu}ij}\right]=\int{d^{3}x}\left[\theta_{ij}-\theta_{ji}\right.\nonumber\\
&&\left.-\frac{e\hbar\kappa}{4mc^{2}}\bar{\psi}\left(\sigma^{0}_{\phantom{\mu}i}F_{0{j}}-\sigma^{0}_{\phantom{\mu}j}F_{0{i}}+\sigma^{k}_{\phantom{\mu}i}F_{k{j}}-\sigma^{k}_{\phantom{\mu}j}F_{k{i}}\right)\psi\right].\nonumber\\
\end{eqnarray}

This equation can be rewritten in a more convenient form as below:
\begin{eqnarray}
d_{t}\boldsymbol{S}+c\boldsymbol{\nabla}S^{0}&=&c\psi^{\dagger}\boldsymbol{\alpha}\times\boldsymbol{\Pi}\psi-\frac{ic\hbar}{2}\boldsymbol{\nabla}\times\boldsymbol{\mathcal{J}}\nonumber\\
&&-\frac{e\hbar\kappa}{4mc}{\psi^{\dagger}}\beta\left(i\boldsymbol{\alpha}\times\boldsymbol{E}-\boldsymbol{\Sigma}\times\boldsymbol{B}\right)\psi,
%
\end{eqnarray}
where $S^{0}=\left(\hbar/2\right)\psi^{\dagger}\gamma^{5}\psi$ and $\mathcal{J}\left(=\psi^{\dagger}\alpha\psi\right)$ is the probability density current.

In the absence of the non-miminal coupling ($\kappa=0$), the standard time evolution of the spin density is recovered. Again, Pauli interaction leads to a torque on the spin density due to the external electric and magnetic fields.

\section{Spin current, spin Hall effect and the Landau levels}\label{spindensitysection}

To model the spintronics phenomenology, one needs to inspect the non-relativisticic regime of the spin density tensor. In what follows, we shall investigate the dynamical equation of the low-relativistic spin density $\boldsymbol{s}=(\hbar/2)\varphi^{\dagger}\boldsymbol{\sigma}\varphi$, where $\boldsymbol{\sigma}$ are the Pauli matrices. 

Hence, to obtain such a limit, one may follow the same procedure discussed in Sec. \ref{Noethersection}. 

Therefore, after algebraic manipulations, the time evolution of the spin density $\boldsymbol{s}$ in the non-relativisticic limit obeys the equation
\begin{eqnarray}\label{spincurrenttimeevolution}
d_{t}\boldsymbol{s}&=&-\boldsymbol{\nabla}\cdot\boldsymbol{\overleftrightarrow{J}}_{s}+\boldsymbol{\mu}_{eff}\times\boldsymbol{B}-\frac{e\hbar^{2}\kappa}{8m^{2}c^{2}}\boldsymbol{\nabla}\times\left(\varphi^{\dagger}\boldsymbol{E}\varphi\right)\nonumber\\
&&-\frac{e\hbar\kappa}{8m^{2}c^{2}}\varphi^{\dagger}\left[\left(\boldsymbol{E}\times\overleftrightarrow{\boldsymbol{\Pi}}\right)\times\boldsymbol{\sigma}\right]\varphi,
\end{eqnarray}
where $\overleftrightarrow{\boldsymbol{\Pi}}\equiv\left[\varphi^{\dagger}\left(\boldsymbol{\Pi}\varphi\right)+\left(\boldsymbol{\Pi}\varphi\right)^{\dagger}\varphi\right]$. This equation tells us that both the electric and magnetic fields contribute to the torque exerted on the spin density. Indeed, it is essential to point out that, in our prescription, the spin precession depends on the applied electric field. It should be contrasted with the situation of the minimal coupling, where only magnetic fields contribute to the time evolution of the spin. 

Furthermore, $\boldsymbol{\overleftrightarrow{J}}_{s}\left(\equiv\mathcal{J}_{ji}\right)$ is the tensor spin current, which is of the form
\begin{eqnarray}\label{generalizedtensorialspincurrent}
\mathcal{J}_{ji}=\frac{i\hbar}{2m}\left[(\nabla_{j}\varphi^{\dagger})s_{i}\varphi-\varphi^{\dagger}s_{i}(\nabla_{j}\varphi)\right]-\frac{e}{mc}{A}_{j}s_{i}.
\end{eqnarray}

It is clear that $\boldsymbol{\boldsymbol{\nabla}}\cdot\boldsymbol{\mathcal{\overleftrightarrow{J}}}$ may be associated to a torque on the 1/2-spin density \cite{Vernes2007}. For $\kappa=0$, one obtains the standard time evolution for the spin density 
\begin{eqnarray}
d_{t}\boldsymbol{s}+\boldsymbol{\boldsymbol{\nabla}}\cdot\boldsymbol{\overleftrightarrow{J}}_{s}=\frac{e}{mc}\boldsymbol{s}\times\boldsymbol{B}, 
\end{eqnarray}
where the term on the rhs is the usual torque exerted by a magnetic field on a spin-1/2 magnetic dipole moment density. This torque is a consequence of the interaction $-\boldsymbol{\mu}\cdot\boldsymbol{B}$ present in the Hamiltonian (\ref{paulitypeequaiton}). 

In absence of external fields, the system above reduces to the continuity equation for the spin current, as expected. On the other hand, the two new torques on the spin degree of freedom that appear on the rhs of Eq. (\ref{spincurrenttimeevolution}) come out as a consequence of the spin-orbit interaction. Let us then discuss the meaning of each of them. The third term on the rhs of Eq. (\ref{spincurrenttimeevolution}) may be associated to a torque due to local changes in the electron charge density, as well as time-varying magnetic fields via the relation $\boldsymbol{\nabla}\times\boldsymbol{E}=-\left(1/c\right)\partial_{t}\boldsymbol{B}$. 
Indeed, such a term has the form required by the spin Hall effect, a phenomenon that takes place when there is an electric field applied along the perpendicular direction of the electrical current propagation. As a consequence, a transverse spin current emerges and gives rise to the so-called {\it intrinsic} spin Hall effect, a phenomenon which is entirely due to the spin-orbit interaction presented in the non-relativistic Hamiltonian of a single electron. Regarding the last term, one notices that a moving electron with a magnetic dipole moment, $\boldsymbol{m}=\left(e\hbar/2mc\right)\varphi^{\dagger}\boldsymbol{\sigma}\varphi$, and velocity, $\boldsymbol{v}$, under the influence of an electric field, $\boldsymbol{E}$, experiences a torque $\boldsymbol{m}\times\boldsymbol{B}'$, where the effective magnetic field turns out to be $\boldsymbol{B}'\approx-\left(e\hbar\kappa/8m^{2}c^{2}\right)\varphi^{\dagger}\left[\left(\boldsymbol{E}\times\overleftrightarrow{\boldsymbol{\Pi}}\right)\times\boldsymbol{\sigma}\right]\varphi$. The $1/2$-factor present in the effective magnetic field $\boldsymbol{B}'$ takes into account the Thomas precession, a kinematical effect that occurs when a charged particle is accelerated due to an applied electric field. 

It should also be highlighted that Eq. (\ref{spincurrenttimeevolution}) may be derived from the local $SU(2)$ gauge symmetry in the Pauli-Schrödinger non-relativisticic theory \cite{Dartora2008,Dartora2010}, as well as by carrying out a Gordon-like decomposition of the total angular momentum current \cite{Wang2006}.

This general approach may be applied to specific systems. In order to evaluate the relevance of the electromagnetic non-minimal coupling in Eq. (\ref{so}), let us  consider 
a static electric field, $\boldsymbol{E}=E_{0}\boldsymbol{\hat{x}}$, together with a constant magnetic field, $\boldsymbol{B}=B_{0}\boldsymbol{\hat{z}}$ . The moving electrons are then confined to move on the plane-$(x,y)$. In such a configuration, the non-relativisticic Hamiltonian (\ref{so}) for the up, $\left(\uparrow\right)$, and down, $\left(\downarrow\right)$, spin, $\sigma_{z}$, assumes the form
\begin{eqnarray}\label{landauhamiltonian}
H_{\uparrow,\downarrow}\!\!&=&\!\!\frac{1}{2m}\left[p_{x}^{2}+\left(p_{y}-\frac{e}{c}B_{0}x\right)^{2}\right]\mp\frac{e\hbar}{2mc}\left(\frac{2+\kappa}{2}\right)B_{0}\nonumber\\
&&\!-\!eE_{0}x\!+\!\frac{e^{2}\hbar^{2}\kappa^{2}E_{0}^{2}}{32m^{3}c^{4}}\!\mp\!\frac{e\hbar\kappa{E_{0}}}{4m^{2}c^{2}}\left(p_{y}\!-\!\frac{e}{c}B_{0}x\right),
\end{eqnarray}
where the Landau gauge $\left(\boldsymbol{A}=xB_{0}\boldsymbol{\hat{y}}\right)$ has been adopted. The Hamiltonian given above commutes with the $z$-component of the spin density $\boldsymbol{s}$, i.e., $\left[H,\sigma_{z}\right]=0$, therefore providing a conserved quantity to compute the spin flux in spin-orbit coupled systems. Furthermore, the system is also translationally invariant along the $y$-direction, which allows us to decompose the energy eingenstates into plane waves, $e^{-ik_{y}y}$, that propagate along the $y$-direction. This decomposition motivates us to use separation of variables for the wave function of the form $\psi_{k_{y}}(x,y)\sim{e^{-ik_{y}y}}f_{k}(x)$. If one applies the Hamiltonian (\ref{landauhamiltonian}) to the aforementioned wave function, then, the energy eingenvalues, $E_{\uparrow,\downarrow}$, turn out to be
\begin{eqnarray}\label{eingenvalues}
E_{\uparrow,\downarrow}\!\!\!&=&\!\!\!\left(n+\frac{1}{2}\right)\!\hbar{w_{B}}\!+\!eE_{0}\!\!\left[l_{B}^{2}k_{y}\!-\!\frac{eE_{0}}{mw_{B}^{2}}\!\pm\!\frac{\hbar\kappa}{8mc}\left(\frac{E_{0}}{B_{0}}\right)\right]\nonumber\\
&&\mp\frac{e\hbar}{2mc}\left(\frac{2+\kappa}{2}\right)B_{0}+\frac{1}{2}mc^{2}\left(\frac{E_{0}}{B_{0}}\right)^{2},
\end{eqnarray}
where $w_{B}=\left(eB_{0}/mc\right)$ is the cyclotron frequency and $l_{B}=\left(\hbar{c}/eB_{0}\right)$ is the strength of the magnetic field. Although the last term of the rhs of the Hamiltonian (\ref{landauhamiltonian}) has a dependence on the momentum $p_{y}$, the energy eingeinvalues (\ref{eingenvalues}) of each level depends linearly on $k_{y}$, in the same way as it occurs in the case of minimal coupling. Furthermore, the splitting between each Landau level is exactly the same as in the standard case, i.e.,  $\triangle{E}_{n}=\hbar{w}_{B}$. On the other hand, the energy splitting between spins is $\triangle=\left(\hbar\kappa/4mc\right)\left(E_{0}/B_{0}\right)-\left(e\hbar/mc\right)B_{0}\left(2+\kappa\right)/2$.

From the energy eingenvalues above, one notices that the first term of the rhs of (\ref{eingenvalues}) denotes a wave packet with momentum $k_{y}$ localized at $x=x_{u,d}$, where 
\begin{eqnarray}
x_{u,d}={l_{B}^{2}}k_{y}-\frac{mc^{2}}{e}\left(\frac{E_{0}}{B_{0}^{2}}\right)\pm\frac{\hbar\kappa}{4mc}\left(\frac{E_{0}}{B_{0}}\right) 
\end{eqnarray}	
is the location of both spin projections, up ($u$) and down ($d$) in the $x$-direction. The latter is simply the kinetic energy of the electron. The other terms can be understood as the potential energy of the wave packet. Note that each spin orientation holds its own Landau level structure, where each level depends linearly on $k_{y}$. Comparing with the Pauli equation, one realizes that the spin-orbit coupling in the Hamiltonian (\ref{landauhamiltonian}) influences the quantum Landau levels, which can be seen by the $\kappa$-contribution in the potential energy (\ref{eingenvalues}).

The wave function, which corresponds to the electron in presence of both electrostatic and magnetostatic fields, is, up to a normalization factor, given by
\begin{eqnarray}
\psi_{n,k_{y}}^{\left(\uparrow,\downarrow	\right)}(x,y)\sim{e^{-ik_{y}y}}H_{n}(\xi)e^{-\xi^{2}/2},
\end{eqnarray}
where $H_{n}$ stand for the standard Hermite polynomials of the harmonic oscillator and $\xi\equiv{x+x_{u,d}}$. Note that the wave function is exponentially localized around $x=x_{u,d}$, but extended over the $y$-direction.


\section{Conclusions and New Prospects}\label{conclusion}

The rôle of the electron spin in electronic systems is a central subject in solid state systems. Typically, in spin-orbit coupled systems, the spin transport is affected by this coupling, which gives rise to interesting phenomena such as the spin Hall effect, for instance. In this vein, we have considered the relativistic Dirac equation coupled in a non-minimal way to an external electromagnetic field. By considering the non-relativistic regime, the spin-orbit interaction shows up. It motivates us to explore the changes in the charge density current, as well as in the spin current. As an immediate application, we have studied the quantum Landau levels and a peculiar effect has appeared as a consequence of the non-minimal coupling in the non-relativistic limit: a spatial splitting between the peak of the wave functions corresponding to the up and down spin components, which may be interpreted as due the appearance of two electronic excitations. 

As a future prospect, we intend to extend our analysis and investigate spin polarization effects and their time evolutions corresponding to the Bargmann-Wigner polarization operator \cite{Bargmann:1948ck,Fradkin1961} in the context of the Dirac equation non-minimally coupled with the electromagnetic field. 

Finally, we would like to stress that there is an intense research in effects related to laser-matter interaction in the dynamics of the plasmas. In this sense, it might be worthy to compute and study exact solutions to the Dirac equation non-minimally coupled to strong electromagnetic fields similar to the situations contemplated in \cite{Raicher:2013cja}. Furthermore, a relativistic generalization of the corresponding many-particle theory may be established \cite{Strange,crepieux2001,mondal2015,mondal2018}, which may lead to a relativistic version of the semi-classical transport theory for the spin Hall effect and for the current-induced switching dynamics, for instance. Also, we point out that the investigation of the Landau levels in the case of neutral particles (the neutron, for example whose electric and magnetic dipole moments are non-vanishing) non-minimally coupled to an external electromagnetic field is another issue we shall pursue, and we shall report on it elsewhere in a forthcoming work. We hope that these interesting features will stimulate further work on the subject. 

\acknowledgments

Rodrigo Turcati  thanks the Physics Department of the Universidade Federal de Santa Catarina for full support. Carlos Andres Bonilla Quintero acknowledges the PCI-DA program of the Observatório Nacional for financial support. This study was financed in part by the Coordenação de Aperfeiçoamento de Pessoal de Nível Superior - Brasil (CAPES) - Finance Code 001.

\end{document}